\documentstyle[12pt,epsf]{article}
\begin{document}
\begin{center}


{\Large \bf Superconducting pairing, and the
collective magnetic excitation in the extended 2-dimensional  
$t-J$ model}\\

\vspace{1.cm}

Oleg Sushkov\\
\vspace{0.5cm}
School of Physics, University of New South Wales,\\
 Sydney 2052, Australia\\
sushkov@newt.phys.unsw.edu.au

\end{center} 

\vspace{1.cm}
 
To investigate the non-Fermi liquid behavior we consider the extended
two-dimensional $t-J$ model which includes additional hopping 
$t^{\prime \prime}$. In the regime $t,J \ll t^{\prime \prime}$
we were able to solve the model analytically. It has a very rich
phase diagram including antiferromagnetic (AF) insulator and
AF strange metal with different kinds of pseudospin-singlet superconducting 
pairings (p,d,g-waves).
We also demonstrate a collective triplet excitation 
with energy below the superconducting gap.\\
 
\vspace{1.cm}
 
\section{Introduction}
 It is now widely accepted that superconductivity of cuprates is
 closely related to their unusual magnetic properties, and
 it is increasingly clear that magnetic pairing is the most realistic
 mechanism of cuprate superconductivity.
 However the mechanism of pairing as well as other unusual properties
 are far from completely understood.
 The problem has been attacked along several directions.
 First we have to mention the empirical or semi-empirical approach
 which allows one to relate different characteristics measured
 experimentally. This approach is to a large extent based on the
 Hubbard model. For a review see article \cite{Chub}.
 In the low energy limit the Hubbard model can be reduced to the
 $t-J$ model. Another approach to cuprates is based on numerical
 studies of the $t-J$ model (see review \cite{Dag}).
 Our studies are also based on this model. We used the
 ordered Neel state at zero doping as a starting point to
 develop the spin-wave theory of pairing \cite{Flam}. The method we used was 
 not fully satisfactory, since it violated spin-rotational symmetry,
 nevertheless it allowed us to calculate from first principles all of the most
 important properties including the critical temperature, the 
spin-wave pseudogap  and the low energy spin triplet excitations \cite{Sus}. 
 
 A sharp collective mode with very low energy has been revealed in YBCO in
 spin polarized inelastic neutron scattering \cite{Ros,Dai,Fong}.
 A number of theoretical explanations have been suggested for this 
 effect \cite{Dem,Sus}, all of these are based on the idea that
 the system is close to AF instability. However all known
 explanations use some uncontrolled approximations and assumptions.
 
In the present work we investigate  close to half filling regime for the  
2D $t-J$ model, where it can be solved analytically without any uncontrolled 
approximations.  It can be done for the region of parameters where long-range 
AF order is  preserved under doping.
This is the regime where non Fermi liquid behavior can be
studied in detail. We analyze the superconducting pairing in this regime
 and consider the spin triplet collective excitation. It is demonstrated
that close to the point of AF instability energy of this excitation is
very small. The excitation exists only at very small momenta.
The idea of this work is somewhat similar to that of our previous 
paper \cite{Damian}, however here we investigate different regime.
 
\section{Hamiltonian and single hole dispersion}
 Let us consider a $t-J-J^{\prime \prime}-V$ model defined by the
 Hamiltonian
 \begin{equation}
 \label{H}
 H=-t\sum_{\langle ij \rangle \sigma} c_{i\sigma}^{\dag}c_{j\sigma}
 -t^{\prime \prime}\sum_{\langle ij_2 \rangle \sigma}
 c_{i\sigma}^{\dag}c_{j_2\sigma}
 + \sum_{\langle ij \rangle} \left[J \left({\bf S}_i{\bf S}_j-
 {1\over 4}n_in_j\right)+Vn_in_j\right].
 \end{equation}
 $c_{i \sigma}^{\dag}$ is the  creation operator of an electron with
 spin $\sigma$ $(\sigma =\uparrow, \downarrow)$ at site $i$
 of the two-dimensional square lattice. The $c_{i \sigma}^{\dag}$ operators
 act in the Hilbert space with no double electron occupancy.
 The $\langle ij \rangle$ represents nearest neighbor sites,
 and $\langle ij_2 \rangle$ represents next next
 nearest sites. The spin operator is ${\bf S}_i={1\over 2}\sum_{\alpha,\beta}
 c_{i \alpha}^{\dag} {\bf \sigma}_{\alpha \beta} c_{i \beta}$,
 and the number density operator is 
 $n_i=\sum_{\sigma}c_{i \sigma}^{\dag}c_{i \sigma}$.
 In addition to the minimal $t-J$ model (see Ref. \cite{Dag}) we have
 introduced additional next next nearest hopping $t^{\prime \prime}$, and 
 Coulomb repulsion $V$ at nearest sites. Note that we do not introduce
 next nearest neighbor hopping $t^{\prime}$ (diagonal) because we do not
 need it for the purposes of this study.

In the paper \cite{Damian} we analyzed the model defined by the Hamiltonian
(\ref{H}) in the limit  $t, t^{\prime \prime} \ll J$.
In the present work we consider limit
\begin{equation}
\label{rel} 
 t,J \ll t^{\prime \prime}.
\end{equation}
It is well known that the $t-J$ model at half filling describes the 
Mott insulator. It is equivalent to the 2D Heisenberg model, and the ground
state of the model has long range AF order. At small doping the
holes are concentrated near the points $(\pm \pi/2,\pm \pi/2)$
where single hole dispersion has minima. In leading approximation
the dispersion is of the form
(we take energy at the minimum as a reference point)
 \begin{eqnarray}
 \label{e}
 &&\epsilon_k= \beta \left(\gamma_{\bf k}^2+(\gamma^-_{\bf k})^2\right),\\
 &&\beta \approx 0.8\times 8t^{\prime \prime}=6.4t^{\prime \prime},
 \nonumber
 \end{eqnarray}
 $\gamma_{\bf k}={1\over 2}(\cos k_x + \cos k_y)$,
 $\gamma_{\bf k}^-={1\over 2}(\cos k_x - \cos k_y)$.
Calculation of the dispersion (\ref{e}) is straightforward because it is 
due to hopping within the same magnetic sublattice. Coefficient $0.8$
appears because of spin quantum fluctuations: $0.8=1-0.2$, where $0.2$ is
the spin flip probability in the Heisenberg model.
 Alon with quasimomentum the hole in the AF background has an additional
quantum number: pseudospin. We denote the hole creation operator by
$h_{{\bf k}\sigma}^{\dag}$, where $\sigma= \pm 1/2$ is pseudospin.
The relation between pseudospin and usual spin is discussed in the 
paper \cite{Saw}.

 We will consider the case of very small doping, $\delta \ll 1$, with 
respect to
 half filling (total filling is $1-\delta$). In this case
 all holes are concentrated in small pockets around the points
 ${\bf k_0}=(\pm \pi/2,\pm \pi/2)$. Single hole dispersion (\ref{e}) can be
 expanded near each of these points
\begin{equation}
\label{ee}
 \epsilon_k= {1\over 2}\beta {\bf p}^2,
\end{equation}
 where ${\bf p}=(p_1,p_2)$ is 
deviation
 from the center of the pocket: ${\bf p}={\bf k-k_0}$, $p_1$ is orthogonal to 
 the face of the magnetic Brillouin zone, and $p_2$ is parallel to the face
(see Fig. 1). The Fermi energy and Fermi momentum for the holes equal
 $ \epsilon_F \approx \frac{1}{2} \pi \beta \delta$, \ 
 $p_F \approx (\pi \delta)^{1/2}$.

\begin{figure}[h]
\vspace{-2pt}
\hspace{-35pt}
\epsfxsize=6cm
\centering\leavevmode\epsfbox{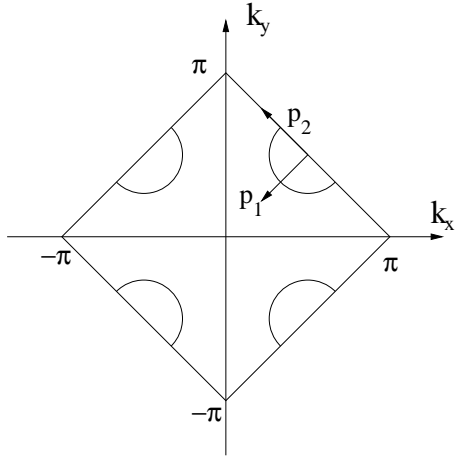}
\vspace{8pt}
\caption{\it {Magnetic Brilloun zone and single hole dispersion}}
\label{Fig1}
\end{figure}

\section{Hole-spin-wave interaction and instability of the Neel state}
 Spin-wave excitations on an AF background are usual spin waves
 with dispersion $\omega_{\bf q}=2J\sqrt{1-\gamma_{\bf q}^2}
 \approx \sqrt{2}Jq$,  at  $q << 1$, see 
 Ref.  \cite{Man} for review. The hole-spin-wave interaction is well known
 (see, e.g. Ref.\cite{Kane})
 \begin{eqnarray}
 \label{hsw}
 && H_{h,sw} = \sum_{\bf k,q} g_{\bf k,q}
     \biggl(
   h_{{\bf k}+{\bf q}\downarrow}^{\dag} h_{{\bf k}\uparrow} \alpha_{\bf q}
   + h_{{\bf k}+{\bf q}\uparrow}^{\dag} h_{{\bf k}\downarrow} \beta_{\bf q}
   + \mbox{H.c.}   \biggr),\\
 && g_{\bf k,q} = 4t\sqrt{2}
  (\gamma_{\bf k} U_{\bf q} + \gamma_{{\bf k}+{\bf q}} V_{\bf q}),\nonumber
 \end{eqnarray}
 where $h_{{\bf k}\sigma}^{\dag}=c_{{\bf k},-\sigma}$ is the hole
 creation operator with pseudospin $\sigma$, 
$\alpha_{\bf q}^{\dag}$ and $\beta_{\bf q}^{\dag}$
 are the spin wave creation operators for $S_z=\mp 1$, and
 $U_{\bf q}=\sqrt{{J\over{\omega_{\bf q}}}+{1\over 2}}$ and\\
 $V_{\bf q}=-sign(\gamma_{\bf q})\sqrt{{J\over{\omega_{\bf q}}}-{1\over 2}}$
 are parameters of the Bogoliubov transformation diagonalizing the spin-wave
 Hamiltonian, see Ref.\cite{Man}. 
Virtual spin wave emission gives a correction to the hole dispersion,
 see Fig.2. However this correction is small 
$\delta \epsilon \sim t^2/t^{\prime \prime}$ and therefore can be neglected
compared with (\ref{e}).

\begin{figure}[h]
\vspace{-2pt}
\hspace{-35pt}
\epsfxsize=4cm
\centering\leavevmode\epsfbox{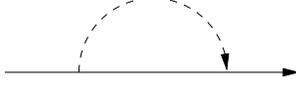}
\vspace{8pt}
\caption{\it {Spin-wave correction to the single hole dispersion.
Solid line corresponds to the hole and dashed line corresponds to the
spin wave.}}
\label{Fig2}
\end{figure}

 To describe renormalization of the spin wave under doping, it is convenient
 to introduce the set of Green's functions \cite{Igar}
 \begin{eqnarray}
 \label{GFs}
 D_{\alpha \alpha}(t,{\bf q})&=&
 -i\langle T[\alpha_{\bf q}(t)\alpha_{\bf q}^{\dag}(0)]\rangle, \\
 D_{\alpha \beta}(t,{\bf q})&=&
 -i\langle T[\alpha_{\bf q}(t)\beta_{\bf -q}(0)]\rangle, \nonumber \\
 D_{\beta \alpha}(t,{\bf q})&=&
 -i\langle T[\beta_{\bf -q}^{\dag}(t)\alpha_{\bf q}^{\dag}(0)]\rangle,
 \nonumber \\
 D_{\beta \beta}(t,{\bf q})&=&
 -i\langle T[\beta_{\bf -q}^{\dag}(t)\beta_{\bf -q}(0)]\rangle.
 \nonumber
 \end{eqnarray}
 In the present work we consider only the long-range dynamics:
 $q \sim k \sim p_F \ll 1$. In this limit all possible polarization
 operators coincide \cite{Sus}
 $P_{\alpha \alpha}(\omega,{\bf q})=P_{\alpha \beta}(\omega,{\bf q})=
 P_{\beta \alpha}(\omega,{\bf q})=P_{\beta \beta}(\omega,{\bf q})=
 \Pi(\omega,{\bf q})$, where $\Pi(\omega,{\bf q})$ is given by the
diagram presented at Fig. 3.

\begin{figure}[h]
\vspace{-2pt}
\hspace{-35pt}
\epsfxsize=4cm
\centering\leavevmode\epsfbox{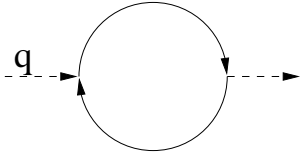}
\vspace{8pt}
\caption{\it {Spin-wave polarization operator.
Solid line corresponds to the hole and dashed line corresponds to the
spin wave.}}
\label{Fig3}
\end{figure}

 For stability of the system the condition (Stoner criterion)
 \begin{equation}
 \label{stab}
 \omega_q + 2 \Pi(0,{\bf q}) > 0
 \end{equation}
 must be fulfilled \cite{Sus1}. Otherwise the Green's functions (\ref{GFs}) 
 would possess poles with imaginary $\omega$. 
 Considering holes as a ``normal Fermi liquid'' \cite{fff} one can easily
 calculate the polarization operator at $q \ll p_F$: 
 $\Pi(0,{\bf q})\approx -4t^2\sqrt{2} q/\pi \beta$,
 Ref. \cite{Sus1}. Relatively weak pairing, which we consider below, does not 
 influence this result. Then the condition of stability can be rewritten as
 \begin{equation}
 \label{stab1}
 \beta =6.4t^{\prime \prime} > {{8t^2}\over{\pi J}}.
 \end{equation}
 To provide stability of the AF order we have to choose 
\begin{equation}
\label{tc}
t^{\prime\prime} > t^{\prime\prime}_c \approx 0.4 t^2/J.
\end{equation}
If $t < J$ or $t \sim J$ the stability condition is automatically
fulfilled since in the present work we consider $t,J \ll t^{\prime \prime}$.
However at $t \gg J$ one can violate the condition (\ref{stab1}).
In this case we will assume that $t^{\prime\prime} > t^{\prime\prime}_c$.
If $t^{\prime\prime}$ is close to $t^{\prime\prime}_c$ it is convenient
to introduce the parameter $\eta$
 \begin{equation}
 \label{eta}
 \eta^2=1-{{8t^2}\over{\pi J\beta}}= 
 (t^{\prime\prime}-t^{\prime\prime}_c)/t^{\prime\prime}
 \end{equation}
as a measure of this closeness.
 The criterion (\ref{stab}) is proportional to this parameter.

\section{Spin-singlet p-wave pairing caused by the short-range attraction}

It is not convenient to consider the superconducting pairing in the
magnetic Brillouin zone with four half-pockets (see Fig. 1). 
Because of this we translate the picture to the shifted zone with two
whole pockets, Fig. 4. We stress that this is question of convenience only,
the representations are absolutely equivalent because of the translational
invariance.

\begin{figure}[h]
\vspace{-2pt}
\hspace{-35pt}
\epsfxsize=6cm
\centering\leavevmode\epsfbox{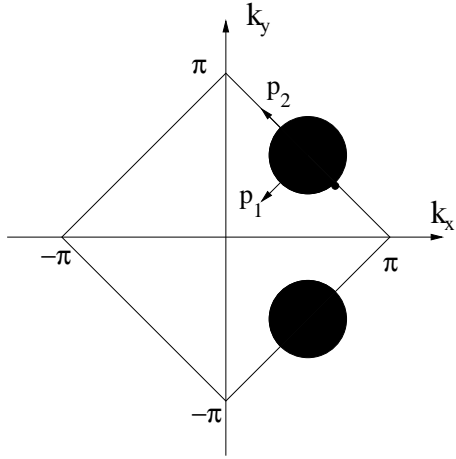}
\vspace{8pt}
\caption{\it {Shifted zone with two whole pockets. Shadowed area
corresponds to occupied hole states.}}
\label{Fig4}
\end{figure}

There are two mechanisms for the superconducting pairing: short-range
attraction and long-range attraction. First we consider the short-range
effect.
Attraction between holes at nearest 
sites (short-range) is due to the reduction in number of missing AF links.
The value of this attraction immediately follows from eq.(\ref{H})
\begin{equation}
\label{U}
U=J\langle{\bf S_i}{\bf S_j}-{1\over 4}+V\rangle\approx -0.58 J+V.
\end{equation}
Strong enough Coulomb repulsion ($V > 0.58 J$) kills this mechanism.
In the momentum representation the interaction (\ref{U}) can be 
rewritten as
\begin{equation}
\label{HU}
H_U=8U\sum_{\bf {k_1,k_2,k_3,k_4}}\gamma_{\bf {k_1-k_3}}
h_{{\bf k_3}\uparrow}^{\dag}h_{{\bf k_4}\downarrow}^{\dag}
h_{{\bf k_2}\downarrow}h_{{\bf k_1}\uparrow}
\delta_{\bf {k_1+k_2,k_3+k_4}}.
\end{equation}
For scattering inside a hole pocket the interaction is practically
momentum independent because 
${\bf {k_1\approx k_2 \approx k_3\approx k_4}}\approx (\pi/2,\pi/2)$,
and hence $\gamma_{\bf {k_1-k_3}}\approx 1$.
Such interaction gives ``s-wave pairing'' with the gap without nodes at
the Fermi surface. The value of the superconducting gap one can easily find 
using the results of papers \cite{Ran,Kuch}. This gives
\begin{equation}
\label{sgap}
\Delta = C t^{\prime \prime}\sqrt{\delta}e^{\pi\beta/4U}
=Ct^{\prime \prime}\sqrt{\delta}
e^{-5t^{\prime \prime}/(0.58J-V)},
\end{equation}
where $C \sim 10$ is some dimensionless constant. The solution is valid
only if $V < 0.58J$, for stronger Coulomb repulsion the pairing
disappears. It is important to stress the peculiar symmetry properties of
the above pairing. This peculiarity comes from the presence of
long-range AF order. As we already mentioned, the gap has no nodes at
the Fermi surface and from this point of view it is ``s-wave pairing''.
However we remind that we have considered the pairing in the shifted zone
and in this zone it is not easy to classify the states by parity.
For well defined parity we have to return to the magnetic Brillouin zone,
so we have to translate the outside parts of the Fermi surface by the inverse
vector of the magnetic lattice ${\bf G}=(\pi,\pi)$, see Fig. 5. 

\begin{figure}[h]
\vspace{-2pt}
\hspace{-35pt}
\epsfxsize=12cm
\centering\leavevmode\epsfbox{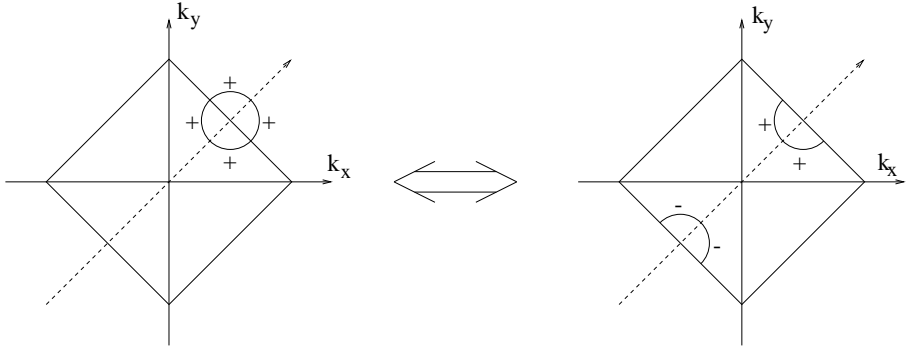}
\vspace{8pt}
\caption{\it {Translation from the shifted zone to the magnetic
Brillouin zone. The superconducting gap has no nodes at the Fermi
surface. The gap changes sign under this
translation.}}
\label{Fig5}
\end{figure}

The point is 
that under such translation the superconducting gap changes the sign as it is 
shown at Fig. 5. This property follows from the fact that the coefficient in 
the interaction (\ref{HU}) changes sign under such translation:
$\gamma_{\bf {k_1-k_3+G}}=-\gamma_{\bf {k_1-k_3}}$ (for details see
paper \cite{Kuch0}).

Thus in reality we have negative parity pairing which is usually called
p-wave. The above consideration was relevant to
the hole pocket centered at $(\pi/2,\pi/2)$. Similar construction
is valid for another pocket centered at $(\pi/2,-\pi/2)$. Existence
of two solutions corresponds to the double degeneracy of the
E-representation of the $C_{4v}$ group. Taking linear combinations
of the single pocket solutions we find two degenerate solutions for
the entire Brillouin zone with lines of nodes
${\bf k_x}=0$ or ${\bf k_y}=0$ well outside the Fermi surface.
We would like to stress that we have considered the spin-singlet (more exactly
pseudospin-singlet) pairing! This situation is very much different
from the usual one when p-wave pairing implies spin triplet. We repeat
that the peculiarity is due to the presence of long-range AF order.

\section{D- and g-wave pairings caused by the long-range attraction}

The long range attraction comes from the spin-wave exchange shown on Fig. 6.
In this exchange the typical spin-wave momenta are 
$q\sim p_F \sim \sqrt{\delta}$,
and hence the typical distances are $r \sim 1/q \sim 1/\sqrt{\delta} \gg 1$.

\begin{figure}[h]
\vspace{-2pt}
\hspace{-35pt}
\epsfxsize=8cm
\centering\leavevmode\epsfbox{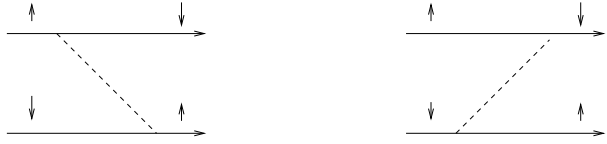}
\vspace{8pt}
\caption{\it {Spin-wave exchange mechanism of attraction.
Solid line corresponds to the hole and dashed line corresponds to the
spin wave. The arrow shows the hole pseudospin.}}
\label{Fig6}
\end{figure}

Similarly to the previous section, it convenient to consider first the pairing
inside a hole pocket, say centered at $(\pi/2,\pi/2)$, see Fig. 4.
This pairing has been considered in detail in our previous work \cite{Flam}.
It has been shown that for the case of ``isotropic'' dispersion (\ref{ee})
the only solution is the one with a single node line in the pocket.
The gap at the Fermi surface  
($\epsilon_F=\frac{1}{2} \pi \beta \delta$) is of the form
 \begin{eqnarray}
 \label{gap}
&& \Delta(\phi)=\Delta_0 \sin \phi,\\
&& \Delta_0= C \epsilon_F \ e^{-\pi J \beta/2t^2}
\approx 10 C t^{\prime\prime}\delta \ e^{- 10 J t^{\prime\prime}/t^2},\nonumber
 \end{eqnarray}
where $\sin \phi =p_2/p_F$, \ \ $p_F^2=p_1^2+p_2^2$,\  and $C\sim 1$ is some 
constant.

 The eqs.(\ref{gap}) describe pairing within a single pocket of the shifted
zone. Translation of this solution to the magnetic Brillouin zone is
shown at Fig. 7. This is absolutely identical to what we did in the previous
section (change of sign at the translation).

\begin{figure}[h]
\vspace{-2pt}
\hspace{-35pt}
\epsfxsize=12cm
\centering\leavevmode\epsfbox{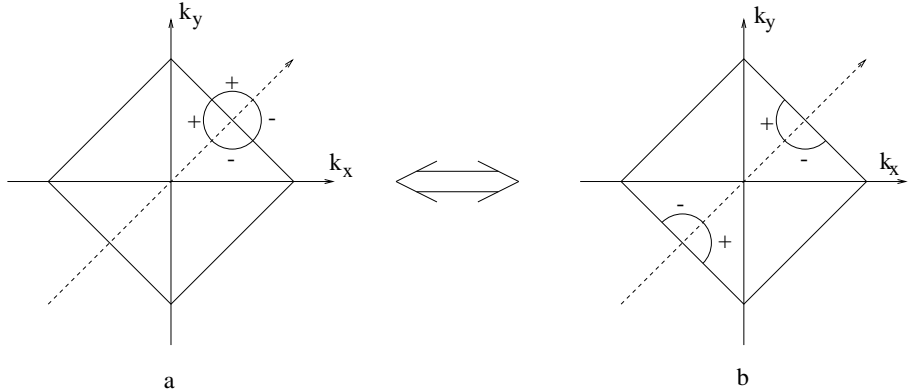}
\vspace{8pt}
\caption{\it {Translation from the shifted zone to the magnetic
Brillouin zone. The superconducting gap has line of nodes. The
gap changes sign under the translation.}}
\label{Fig7}
\end{figure}

There are effectively two pockets in the  Brillouin zone, see Fig. 4. Taking 
symmetric and antisymmetric combinations between the pockets,
 we get the d- and g-wave 
pairings respectively. The symmetries of the corresponding superconducting
gaps are shown at Fig. 8. It is clear that the d-wave belongs to the 
$B_1$ representation of the $C_{4v}$ group and the g-wave belongs to the $A_2$
representation.

\begin{figure}[h]
\vspace{-2pt}
\hspace{-35pt}
\epsfxsize=12cm
\centering\leavevmode\epsfbox{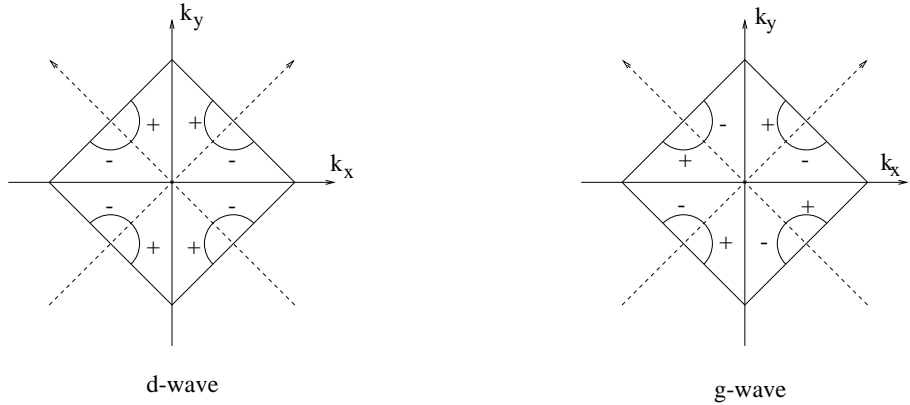}
\vspace{8pt}
\caption{\it {Symmetry of the superconducting gap corresponding to
the d- and g-wave pairings.}}
\label{Fig8}
\end{figure}

Both solutions originate from (\ref{gap}), therefore they
are close in energy. Nevertheless the constant $C$ in eq. (\ref{gap})
is smaller for the g-wave. This is the price for additional lines of nodes
(${\bf k_x}=0$ and ${\bf k_y}=0$). The above consideration did not include
short range interaction (\ref{HU}). This is absolutely correct for g-wave
pairing which is not sensitive to the interaction (\ref{HU}) at all.
However the d-wave is sensitive.
Therefore at $V < 0.58 J$ the d-wave pairing is enhanced
because of (\ref{HU}), while, on the contrary, at larger Coulomb repulsion 
$V > 0.58 J$ the d-wave is suppressed and can even disappear.
To avoid misunderstanding we stress that in the limit under
consideration ($t^{\prime\prime} \gg t,J$) the short range interaction
(\ref{HU}) is too weak (even at $V=0$) to produce d-wave pairing
without spin-wave exchange. However the short-range interaction  influences
the dimensionless constant $C$ (see eq. (\ref{gap})) which arises in
spin-wave exchange mechanism.

\section{The phase diagram}
The phase diagram of the model under consideration is given on Fig. 9.
To be specific we present the case of the not too strong Coulomb repulsion 
at the nearest sites: $V < 0.58J$. At stronger $V$ the p-wave
superconductivity disappears, see eq. (\ref{sgap}). 
Comparing eqs. (\ref{sgap}) and (\ref{gap}) we
see that the p-wave pairing is stronger at $t < t_c$, while at
$t > t_c$ the d-g-wave pairing dominates. At $V=0$ the critical value
is $t_c \approx J$.
In the p-wave phase the gap, as well as the critical temperature,
is proportinal to square  root of the hole concentration:
$\Delta \sim T_c \propto \sqrt{\delta}$. But in the d-g-wave phase
they are proportional to the first power of concentration:
$\Delta \sim T_c \propto \delta$

\begin{figure}[h]
\vspace{-2pt}
\hspace{-35pt}
\epsfxsize=10cm
\centering\leavevmode\epsfbox{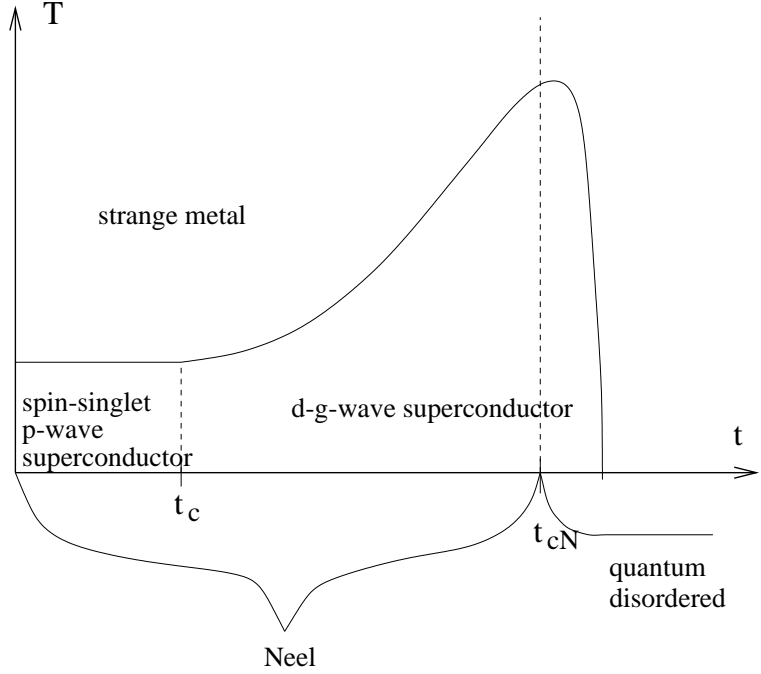}
\vspace{8pt}
\caption{\it {The phase diagram of the extended 2D $t-t^{\prime\prime}-J-V$ 
model. $t_c$ is transition point from the p-wave to the d-g-wave
superconductor. $t_{cN}$ is transition point from the Neel state to the
spin quantum disordered state.}}
\label{Fig9}
\end{figure}

According to eq. (\ref{tc}) at 
$t < t_{cN} \approx 1.6 \sqrt{t^{\prime\prime}J}$, the  long range AF order 
at zero temperature is preserved under doping, so we have coexistence of the
superconductivity and the Neel order. At
$t > t_{cN}$ the Neel order is destroyed by the
doping and one gets a transition into the quantum disordered phase.
However as soon as the magnetic correlation length is larger than the
superconducting correlation length the mechanism of pairing is valid
and one still has the d-g-wave superconductor.
At a temperature higher than the critical one the system behaves as a metal
with very strong scattering of mobile holes on spin-wave excitations.
Following the tradition we call this state ``strange metal''.

\section{The spin-wave collective excitation}
We will see that the spin-wave collective excitation has nontrivial
behaviour in the vicinity of the quantum phase transition from the
Neel to the disordered phase. Therefore we study this excitation
only in the d-g-wave superconducting phase at $T=0$. The
 energy spectrum and Bogoliubov parameters are given by the usual BCS
 formulas
 \begin{eqnarray}
 \label{BCS}
 E_{\bf k}&=&\sqrt{(\epsilon_k-\epsilon_F)^2+\Delta_{\bf k}^2},\\
 u_{\bf k}^2, v_{\bf k}^2
 &=&{1\over 2}\left(1\pm{{\epsilon_{\bf k}-\epsilon_F}\over
 {E_{\bf k}}}\right)\nonumber
 \end{eqnarray}
with gap $\Delta_{\bf k}$ from eq. (\ref{gap}).
The spin wave  polarization operator due to mobile holes is given by
diagram on Fig. 3 plus a similar diagram with anomalous fermionic Green's
 functions. Straightforward calculation gives
 (see e.g. Ref.\cite{Sus})
 \begin{equation}
 \label{pol}
 \Pi(\omega,{\bf q})=\sum_{\bf k,k_0}g_{\bf k_0 q}^2{{2(E_{\bf k}+E_{\bf k+q})}
 \over{\omega^2-(E_{\bf k}+E_{\bf k+q})^2}}
 \left(u_{\bf k}^2v_{\bf k+q}^2+u_{\bf k}v_{\bf k}u_{\bf k+q}v_{\bf k+q}
 \right).
 \end{equation}
This equation includes summation over
 pockets ${\bf k_0}=(\pi/2,\pm\pi/2)$. In these pockets
 the vertex (\ref{hsw}) is 
$g_{\bf k_0,q} \approx 2^{5/4}t(q_x \pm q_y)/\sqrt{q}$.
 Let us consider the case of very small momenta and frequencies:
 $v_Fq < \Delta_0$, and $\omega < \Delta_0$.
 In this limit one can put $q=0$ in eq. (\ref{pol}) everywhere except at the
 vertex and therefore the polarization operator can be evaluated
 analytically
 \begin{equation}
 \label{pol1}
 \Pi(\omega,{\bf q})=-{{4t^2\omega_{\bf q}}\over{\pi J \beta}}
 \left(1+i{{\pi\omega}\over{8\Delta_0}}\right)
 \end{equation}
 Note that the the imaginary part is nonzero even at $\omega < 2\Delta_0$
 because the gap (\ref{gap}) has a line of nodes.
 Any of the Green's functions (\ref{GFs}) have a denominator
 $\omega^2-\omega_{\bf q}^2-2\omega_{\bf q}\Pi(\omega,{\bf q})$,
 see e.g. Refs. \cite{Igar,Sus}. The zero of this denominator gives the energy
 and width of the spin-triplet collective excitation. Using eqs.(\ref{pol1})
 and (\ref{eta}) we find
 \begin{eqnarray}
 \label{o}
 o_{\bf q}&=&\eta \omega_{\bf q},\\
 \Gamma_{\bf q}&=&{{\pi}\over{8}}{{1-\eta^2}\over{\eta}}
{{\omega_{\bf q}}\over{\Delta_0}}o_{\bf q}.\nonumber 
 \end{eqnarray}
 In essence this is the renormalized spin-wave. Far from the point of
AF instability the parameter $\eta \approx 1$,  therefore
the renormalization is relatively weak and the decay width is small.
The situation is different when approaching the point of instability
$t\to t_{cN}\approx 1.6\sqrt{t^{\prime\prime}/J}$. Here, according to 
eq. (\ref{eta}),
$\eta \to 0$ and therefore the energy of the renormalized spin wave
is much smaller than the energy of the bare spin-wave, 
$o_{\bf q}/\omega_{\bf q}=\eta \ll 1$. Moreover this collective excitation 
exists as a narrow peak only at very small $q$, when 
\begin{equation}
\label{lim}
\pi \omega_{\bf q}/8\eta\Delta_0 <1.
\end{equation}
At higher $q$ the width is larger than its frequency because of decay to 
particle-hole excitations. We stress that the closer to the point of 
instability, the smaller is $\eta$, and
therefore the smaller is the region of $q$ where the excitation exists.

\section{Conclusions}

 We have considered a close to half filling 
 $t-t^{\prime \prime}-J-V$ model at $t^{\prime \prime} \gg t,J$.
 We restrict our consideration to the case of small doping $\delta \ll 1$.
 It is demonstrated that at 
$t < t_{cN} \approx 1.6\sqrt{t^{\prime\prime} J}$
the Neel order is preserved under the doping, and at 
$t > t_{cN}$ the order is destroyed and the system undergoes a transition to 
the  quantum spin disordered phase, see phase diagram at Fig. 9.

If the hole-hole Coulomb repulsion at nearest sites is not too strong
($V < 0.58 J$), then at small $t$ the model has psedospin-singlet
p-wave superconductivity. As $t$ increases, at the some point $t_c$
(at $V=0$ the critical point is $t_c\approx J$) the system
undergoes a phase transition from the p-wave
to the d-g-wave superconductor, see Fig. 9.
Which state is realized (d- or g-wave) crucially depends on the
Coulomb repulsion $V$. If $V$ is small the d-wave is preferable,
while at larger $V$ the g-wave superconductivity is realized.

 In the Neel state we found the collective spin triplet excitation
(renormalized spin wave). In the vicinity of the quantum phase
transition to the spin disordered state the excitation
exists as a narrow mode only at very small momenta and its energy is
 substantially below the energy of the bare spin wave.
 

 

\begin{thebibliography}{99}
\bibitem{fff} ``Normal Fermi liquid'' is a somewhat misleading term
because we consider a system which is not a normal Fermi liquid from
the common point of view. In this case ``Normal Fermim liquid'' means
that we consider an ideal gas of Fermions with dispersion (\ref{e}).
At zero temperature all states inside the semicircles at the Fig. 1
are occupied.
\end{thebibliography}
 \end{document}